\documentclass[aps,pra,twocolumn,groupedaddress]{revtex4-1}

\usepackage{amsmath}
\usepackage{bm}
\usepackage{graphics}
\usepackage{times}
\usepackage{graphicx}
\usepackage{xcolor}
\usepackage{array,multirow,graphicx}
\usepackage{multirow,hhline,graphicx,array}
\usepackage{rotating}
\usepackage[markup=underlined]{changes}

\begin{document}

\title{Multi-sideband interference structures observed via high-order
   photon-induced continuum-continuum transitions in argon}

\author{
D Bharti$^{1}$, H~Srinivas$^{1}$, F~Shobeiry$^{1}$, 
K R Hamilton$^{2}$, R~Moshammer$^{1}$, T~Pfeifer$^{1}$, K~Bartschat$^{2}$, 
and A~Harth$^{1,3}$
} 

\email{bharti@mpi-hd.mpg.de}
\email{Anne.Harth@hs-aalen.de}

\affiliation{$^1$Max-Planck-Institute for Nuclear Physics, D-69117 Heidelberg, Germany}
\affiliation{$^2$Department of Physics and Astronomy, Drake University, Des Moines, IA 50311, USA}
\affiliation{$^3$Department of Optics and Mechatronics, Hochschule Aalen, D-73430 Aalen, Germany}

\date{\today}

\begin{abstract}
We report a joint experimental and theoretical study of a three-sideband \hbox{(3-SB)} modification of the
``reconstruction of attosecond beating by interference of two-photon transitions'' \hbox{(RABBIT)} setup.  
The \hbox{3-SB} RABBIT scheme makes it possible to investigate phases resulting from interference 
between transitions of different orders in the continuum.
Furthermore, the strength of this method is its ability 
to focus on the atomic phases only, independent of a 
chirp in the harmonics, by comparing the RABBIT phases extracted from specific SB groups formed by two adjacent harmonics.
We verify earlier predictions that the phases and the corresponding time delays in the three SBs extracted from 
angle-integrated measurements become similar with increasing photoelectron energy. 
A variation in the angle-dependence of the RABBIT phases in the three SBs results from the distinct Wigner and 
continuum-continuum coupling phases associated with the individual angular-momentum channels.  A qualitative explanation
of this dependence is attempted by invoking a propensity rule. 
Comparison between the experimental data and predictions from an $R$-matrix (close-coupling) 
with time dependence calculation shows qualitative agreement in most of the observed trends.
\end{abstract}

\maketitle

\section{Introduction} 
The reconstruction of attosecond beating by interference of two-photon transitions \hbox{(RABBIT)} 
is a widely employed technique to measure attosecond time delays in photo\-ionization processes \cite{Paul2001,Muller2002,Klunder2011}. 
The extraction of time information from the RABBIT measurements usually involves retrieving 
atomic phases encoded in the delay-dependent modulation of the sideband (SB) yield. 
These SBs are traditionally formed in the photo\-electron spectrum by the interaction 
of two photons (one pump, one probe) with the target. Spectral harmonics from an attosecond pulse train (the pump photons)
form discrete photo\-electron signal peaks. The presence of a time-delayed infrared field (the probe photon)
then creates a signal in between these main peaks that oscillates with the time delay. 
The so retrieved atomic phase ($\Delta\phi^{at}$) from the RABBIT measurement can be 
separated into a single-photon ionization contribution ($\Delta\eta$, Wigner phase \cite{PhysRev.98.145}) 
and a continuum-continuum (cc) coupling phase ($\Delta\phi^{cc}$) by applying 
an ``asymptotic approximation'' \cite{Dahlstr_m_2012,Dahlstr_m_2014,RevModPhys.87.765}. 

Variations of the RABBIT scheme, such as \hbox{0-SB}, \hbox{1-SB}, and \hbox{2-SB}, have been utilized to study 
dipole transition phases and attosecond pulse shaping \cite{Loriot_2017,PhysRevA.104.043113, Maroju2020all}. 
As the name suggests, in a \hbox{3-SB} \hbox{RABBIT} scheme, three SBs are formed between 
two consecutive main photoelectron peaks \cite{Harth2019,Bharti2021}. 
The delay-dependent oscillation in the photo\-electron signal of these three SBs requires more than one transition in the continuum, i.e., the absorption or emission of several probe photons. 
For a hydrogenic system, we recently~\cite{Bharti2021} extended the asymptotic approximation to a decomposition scheme, which expands the phase of the $N^{\rm th}$-order dipole matrix element ${\cal M}^{(N)}$, describing the absorption of an ionizing extreme ultra\-violet (XUV) photon followed by $N\!-\!1$ infra\-red (IR) photon exchange in the continuum, into a sum of the Wigner phase and $N\!-\!1$ cc phases.


For atomic hydrogen, where numerical calculations with high accuracy can be carried out by 
solving the time-dependent Schr\"odinger equation (TDSE) directly, we verified that the 
decomposition approximation explains the \hbox{RABBIT} phases in all three SBs  
qualitatively~\cite{Bharti2021}.  
As expected, its accuracy improves with increasing energy of the emitted photo\-electron.
On the other hand, assuming $\Delta\phi^{cc}$ to be independent of the orbital angular 
momenta of the continuum states leads to deviations from the analytical prediction, 
particularly in the lower and the higher SB of the triplet at low kinetic energies. 

Even though starting with a $3p$ electron still limits the information 
that can be extracted due to the combined effect of at least {\it two}  Wigner and the cc phases, 
we decided to
perform the present proof-of-principle 
study on argon due to its experimental advantages, including a significantly lower 
ionization potential than helium, which may be a viable alternative
to atomic hydrogen due to its quasi-one-electron character, as long as one of the 
electrons remains in the $1s$ orbital, i.e., away from doubly-excited resonance states.
In argon, the intermediate orbital angular momentum after the XUV step is \hbox{$\lambda \!=\! 0~\rm or~2$}, 
while \hbox{$\lambda \!=\! 1$} in helium.
For the latter target, as for atomic hydrogen, the dependence on the Wigner phase would drop out, and the \hbox{3-SB} 
setup would provide direct access to the phase associated with 
higher-order cc transitions~\cite{Harth2019,Bharti2021}. 
Nevertheless, a significant strength of our current setup already lies in the fact that the results within each group 
are {\it independent of any chirp} in the XUV pulse, because the XUV harmonic pair 
is common to all three SBs. 

This paper is organized as follows. We begin with a brief review of the basic idea behind
the \hbox{3-SB} setup in Sec.~\ref{sec:basic}. This is followed by a description of the experimental 
apparatus in Sec.~\ref{sec:experiment} and the accompanying theoretical \hbox{$R$-matrix} (close-coupling) with 
time dependence (RMT) approach in Sec.~\ref{sec:theory}.
In section~\ref{sec:results}, we first show angle-integrated data
(Sec.~\ref{subsec:Angle-integrated})
before focusing on the angle-dependence of the \hbox{RABBIT} phases in the three SBs of each individual group in 
Sec.~\ref{subsec:Angle-differential}.  We finish with a summary and an outlook in Sec.~\ref{sec:outlook}.

\section{The 3-SB Scheme}\label{sec:basic}
 In this section, we briefly review the 3-SB scheme introduced in~\cite{Harth2019} and the analytical 
treatment presented in~\cite{Bharti2021} as applied to the \hbox{3-SB} RABBIT experiment.

\begin{figure}[h]
\includegraphics[width=0.9\columnwidth]{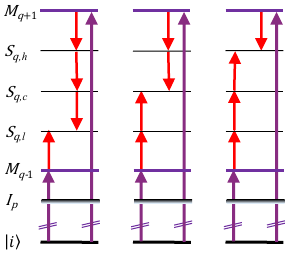}
\caption{3-SB \hbox{RABBIT} scheme. $M_{q-1}$ and $M_{q+1}$ label the main 
photoelectron peaks created directly by the odd harmonics ($H_{q-1}$ and $H_{q+1}$) 
of the frequency-doubled fundamental probe frequency in the XUV pulse, while $S_{q,l}$, $S_{q,c}$, 
and $S_{q,h}$ are the lower, central, and higher SBs, respectively.  These SBs 
are formed by emission or absorption of probe photons by the quasi-free photo\-electrons.
$|i\rangle$ denotes the initial state and $I_p$ is the ionization potential.}
\label{fig:scheme}
\end{figure}

\begin{figure*}
\includegraphics[width=0.99\linewidth]{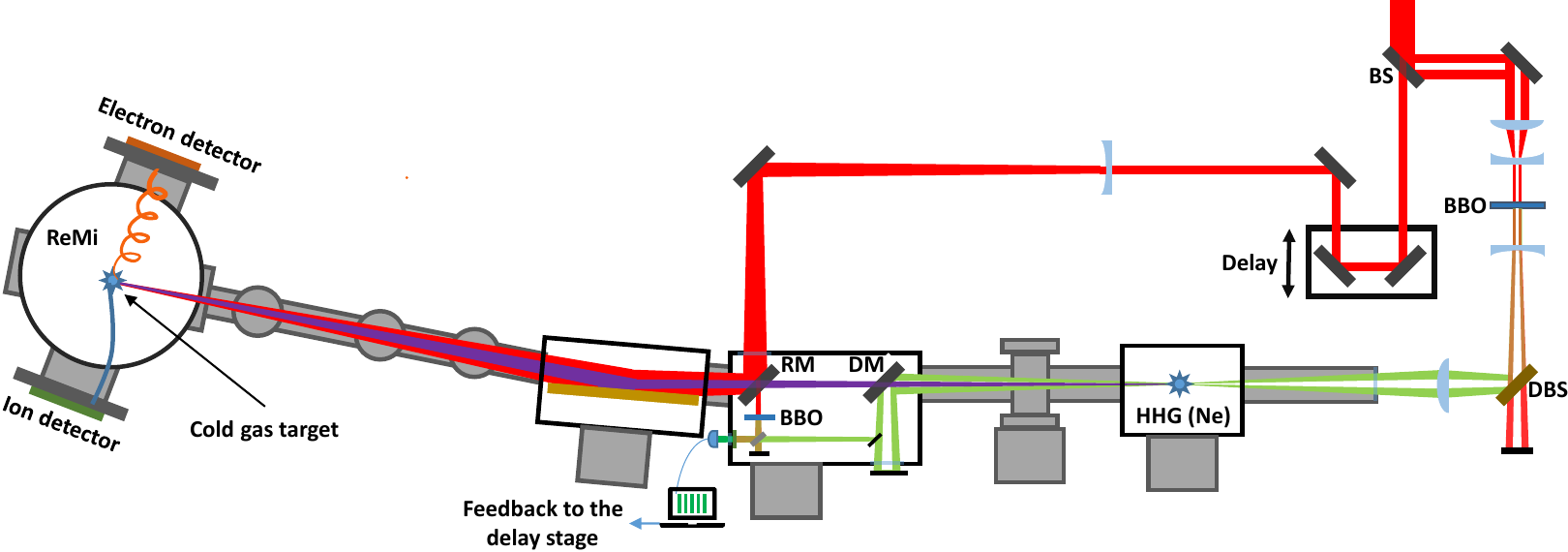}
\caption{Experimental setup. A holey mirror (BS) splits the linearly polarized laser 
beam between the two arms of the inter\-ferometer. In the pump arm, the HHG process is 
driven by the second harmonic of the laser beam. The generated XUV and the fundamental 
probe beam are recombined and focused onto a supersonic gas jet of Argon. The 
interferometer is stabilized by tracking the movement of the fringes from the pump and the probe beams.}
\label{fig:setup}
\end{figure*}
Figure~\ref{fig:scheme} illustrates only the two most dominant transition paths 
for each SB contributing to the oscillation in their respective yields.
The lowest-order transition dominates the yield, but its modulation requires 
inter\-ference between at least two distinct paths leading to the same energy. 
This involves two different XUV harmonics that are aided by absorption or emission of 
near-infra\-red (NIR) photons. 
For the lower ($l$) and higher ($h$) SBs, $S_l$ and $S_h$, the most important interfering 
paths are of 2$^{\rm nd}$ (one harmonic and one NIR) and 4$^{\rm th}$ (one harmonic and three NIR) order, 
which results in a weak modulation of the yield. The lowest-order terms contributing to the build-up 
of the central ($c$) SB, $S_c$, are both of 3$^{\rm rd}$ order (one harmonic and two NIR).  Consequently, 
interference between them exhibits the delay-dependent oscillation most clearly.

Mathematically, the angle-integrated yield in the three SBs, considering only two prominent transition 
paths, can be written as
\begin{subequations}
\begin{align}
S_{q,l}\!&\propto \!\! \sum_{\ell,m} \!\left|\!\left( \tilde{E}_{q+1}\tilde{E}^{*3}_{\omega} 
{\cal M}^{(4,e)}_{\ell,m}(k_{l,q}) \!+\! \tilde{E}_{q-1}\tilde{E}_{\omega} {\cal M}^{(2,a)}_{\ell,m}(k_{l,q}) \right) \!\right|^{2} \nonumber\\
&= I^l_0 + \sum_{\ell,m} I^l_{\ell,m} \, \cos(4\,\omega\tau -\Delta\phi_{\Omega}^{q}-\Delta\phi_{\ell,m}^{l,at})  \nonumber\\ 
&= I^l_0 + I^l_1 \, \cos(4\,\omega\tau -\phi^l_{R}+\pi);\\ 
S_{q,c}\!&\propto \!\! \sum_{\ell,m} \!\left|\!\left(\tilde{E}_{q+1}\tilde{E}^{*2}_{\omega} 
{\cal M}^{(3,e)}_{\ell,m}(k_{c,q}) \!+\! \tilde{E}_{q-1}\tilde{E}^2_{\omega} {\cal M}^{(3,a)}_{\ell,m}(k_{c,q}) \right) \!\right|^{2} \nonumber\\
&= I^c_0 + \sum_{\ell,m} I^c_{\ell,m} \, \cos(4\,\omega\tau -\Delta\phi_{\Omega}^{q}-\Delta\phi_{\ell,m}^{c,at})  \nonumber\\ 
& = I^c_0 + I^c_1 \, \cos(4\,\omega\tau -\phi^c_{R});\\ 
S_{q,h}\!&\propto \!\! \sum_{\ell,m} \!\left|\!\left(\tilde{E}_{q+1}\tilde{E}^{*}_{\omega} 
{\cal M}^{(2,e)}_{\ell,m}(k_{h,q}) \!+\! \tilde{E}_{q-1}\tilde{E}^3_{\omega} {\cal M}^{(4,a)}_{\ell,m}(k_{h,q}) \right) \!\right|^{2} \nonumber\\
&= I^h_0 + \sum_{\ell,m} I^h_{\ell,m} \, \cos(4\,\omega\tau -\Delta\phi_{\Omega}^{q}-\Delta\phi_{\ell,m}^{h,at})  \nonumber\\ 
& = I^h_0 + I^h_1 \, \cos(4\,\omega\tau -\phi^h_{R}+\pi) 
\end{align}
\label{eq: SB osc c}
\end{subequations}
\noindent Here $q$ labels the SB group, while $k_{l,q}$, $k_{c,q}$, and $k_{h,q}$ denote  
the final linear momenta of the ejected electron in the lower, central, and higher side\-bands in each group.
The subscript~$\ell$ denotes one of generally several allowed orbital angular momenta  of the ejected electron in the final state and~$m$ labels the magnetic quantum number, which can be $0$ or $\pm 1$ for the electron starting in the $3p$ sub\-shell.
Note that $m$ is a conserved quantity for all orders~$n$ of the transition matrix element ${\cal M}^{(n)}_{\ell,m}$ due to our use of linearly polarized light.

Furthermore, \hbox{$\tilde{E}_{\Omega}=E_{\Omega}{\rm e}^{{i}\, \phi_\Omega}$} and \hbox{$\tilde{E}_\omega=E_\omega{\rm e}^{{i}\, \omega \tau}$}  (for absorption) are the complex electric-field amplitudes of the  co-linearly polarized XUV-pump ($\Omega$) and NIR-probe ($\omega$) pulses, respectively. 
$\Delta\phi_{\ell,m}^{at} =\mathrm{arg}[{\cal M}^{(a)}_{\ell,m}{\cal M}^{*(e)}_{\ell,m}]$ 
is the phase difference between the two matrix elements and $a(e)$ denotes the pathway 
involving absorption (emission) of the probe photons. 
Finally, $\Delta\phi_{\Omega}^{q}$ is the spectral phase difference (XUV chirp) of two neighbouring harmonics.


As seen from Eqs.~(1), the yield of each SB is separated into an average part $I_0$ and another term $I_1$ that oscillates at $4\,\omega$ with the delay. 
As discussed in~\cite{Bharti2021}, 
every dipole transition also adds a phase of~$\pi/2$. Since the two dominant interfering terms 
in $S_l$ and $S_h$ are of different orders (2$^{\rm nd}$ and 4$^{\rm th}$), this leads to an 
additional $\pi$ phase in $S_l$ and $S_h$ relative to $S_c$, where both interfering terms 
are of the same (3$^{\rm rd}$) order. 

The RABBIT phase ($\phi_{R}$) includes the spectral phase difference of the two harmonics and 
the channel-resolved atomic phases weighted according to their transition amplitudes. 
It is a complex inverse trigonometric function involving many parameters and hence is best determined
by fitting the signal to the known analytic form given above.
Since the three SBs involve 
the same pair of harmonics, the contribution of the XUV group delay (i.e., the chirp) to the oscillation phase is the 
same in all three SBs. This is a key advantage of the \hbox{3-SB method}, since it removes the 
influence of the XUV chirp when we compare the phases of the three SBs only within a particular group.

\section{Experimental Setup}\label{sec:experiment}
Figure~\ref{fig:setup} shows the schematic design of our \hbox{3-SB} \hbox{RABBIT} experimental setup. 
A commercial fiber-based laser delivers pulses with a duration of approximately 50~fs (FWHM) at a 
49~kHz repetition rate with a pulse energy of 1.2~mJ and a center wavelength of 1030~nm.
This pulse is split into two parts using a holey mirror (BS) that reflects $\approx 85 \% $ of 
the incoming beam in the pump arm, while the rest passes through the hole into the probe arm. 
The beam size of the reflected donut beam in the pump arm is reduced by a pair of lenses and 
passed through a 0.5~mm thick BBO crystal to double its frequency. 

The conversion efficiency for the Second-Harmonic Generation (SHG) by the BBO crystal is \hbox{$25-30\,\%$.} 
A dichroic beam-splitter (DBS) filters out the fundamental beam, and a lens with a focal 
length of 12~cm focuses the second harmonic beam inside a vacuum chamber to a focal spot 
of $30-40~\mu\,$m on a jet of neon gas, which results in an XUV frequency comb through 
high-harmonic generation (HHG). The gas nozzle has a diameter of $100~\mu\,$m and is operated 
at a backing pressure of 1.2~bar with a chamber pressure of $5 \times 10^{-3}$~mbar. 
The generated XUV beam is spatially separated from the annular second harmonic with the help of an 
additional holey dumping mirror (DM). The residual second harmonic passed through the dumping mirror is weak and does not generate any visible sidebands. 
The beam in the probe arm goes through a retro-reflector mounted on a piezo\-electric-translation 
stage that offers a step-resolution of 5~nm with closed-loop position control. 
Another holey mirror (RM) recombines the NIR (probe) and XUV (pump) beams, which are then focused 
inside a reaction microscope (ReMi) on a cold gas jet of argon. 
The ReMi enables coincident detection and the reconstruction of the three-dimensional momenta of the 
ions and electrons created during the photo\-ionization process~\cite{Ullrich_2003}.
The inter\-ferometer was actively stabilized~\cite{Srinivas:22} to achieve a stability of 
\hbox{$\approx 40\,$}atto\-seconds 
over a data acquisition time of seven hours. 
The stability of the inter\-ferometer was critical for the successful realization of the
\hbox{3-SB} scheme since the oscillation period was just $850\,$atto\-seconds.

\smallskip
\section{Theoretical Approach}\label{sec:theory}
In the theoretical part of this study, we employ the general \hbox{$R$-matrix} with time dependence (RMT)
method~\cite{BROWN2020107062} to generate theoretical predictions for comparison with our experimental data.
In order to calculate the necessary time-independent basis functions and dipole matrix elements, we set up the \hbox{2-state} 
non\-relativistic model introduced by Burke and Taylor~\cite{BurkeTaylor1975} 
to treat the steady-state standard photo\-ionization process.  
In this model, multi-configuration expansions for the initial $(3s^23p^6)^1S$ 
bound state and the two coupled final ionic states $(3s^2 3p^5)^2P$ and $(3s 3p^6)^2S$ were employed.  
We checked that the photo\-ionization cross sections at the photon energies 
corresponding to the various HHG lines was reproduced properly (in agreement 
with Burke and Taylor~\cite{BurkeTaylor1975} as well as  experiment~\cite{MarrWest1976,SAMSON2002265}) by our RMT model.

\begin{figure*}[t]
\includegraphics[width=0.99\linewidth]{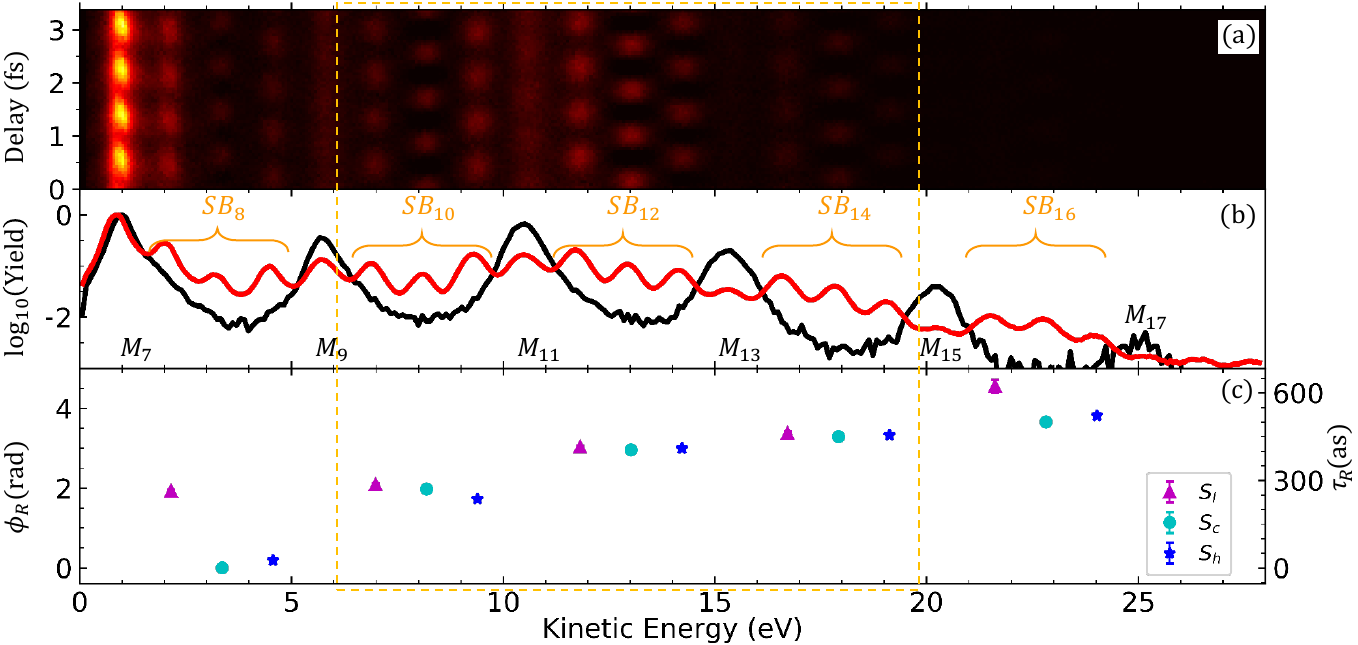} 
\caption{3-SB RABBIT trace~(a), normalized photo\-electron spectra generated with the XUV 
pulse only (dark) and during the RABBIT measurement integrated over the delays (lighter)~(b), 
and \hbox{RABBIT} phases extracted from all three sidebands~(c). Note that the $\pi$ phase 
difference between $S_c$ and ($S_l,S_h$), which is clearly seen in the position of 
the maxima in panel~(a), has been removed for better visibility in panel~(c).
The error bars from the fitting procedure are generally smaller than the symbol size and hence hardly visible.
The dashed box from about 6~eV to 21~eV indicates the sideband groups that we concentrate
on for the angle-differential cases.}
\label{fig:expt-results}
\end{figure*} 

The probe-pulse duration was chosen as about twice the length of the XUV pulse. 
We emphasize that the present calculation was meant as a supplement to the current experiment, with the hope of providing  
additional qualitative insights rather than quantitative agreement, which would require much more detailed information
about the actual pulses than what was available.  We purposely employed significantly lower NIR peak 
intensities ($10^{11}\,$W/cm$^2$) than in 
the experiment ($\approx 6 \times 10^{11}\,$W/cm$^2$).  This reduced the number of partial 
waves needed to obtain converged results, diminished potential distortions, and thus made it easier to interpret the spectra.

Specifically, we performed calculations for 11 delays in multiples of
0.05 NIR periods.  
For each delay, we needed about 5 hours on 23 nodes using all 56 available cores per node 
on the \hbox{Frontera} supercomputer hosted at the Texas Advanced \hbox{Computing} Center (TACC)~\cite{frontera}.

\section{Results and Discussion}\label{sec:results}
Below we present our results.  We start with the angle-integrated setup in Sec.~\ref{subsec:Angle-integrated} 
before going into further detail with angle-resolved measurements and calculations in Sec.~\ref{subsec:Angle-differential}. 

\subsection{Angle-integrated RABBIT phases}\label{subsec:Angle-integrated}

\begin{table*}
\includegraphics[width=0.9\linewidth]{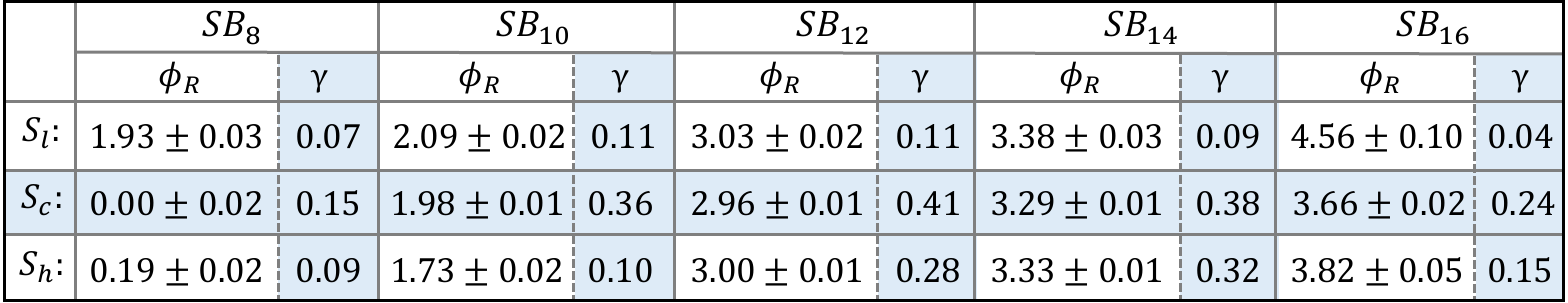} 
\caption{RABBIT phase extracted from the fitting procedure and the contrast~$\gamma$ of the oscillation.}
\label{tab:Tab1}
\end{table*} 

\begin{figure}
\includegraphics[width=0.80\linewidth]{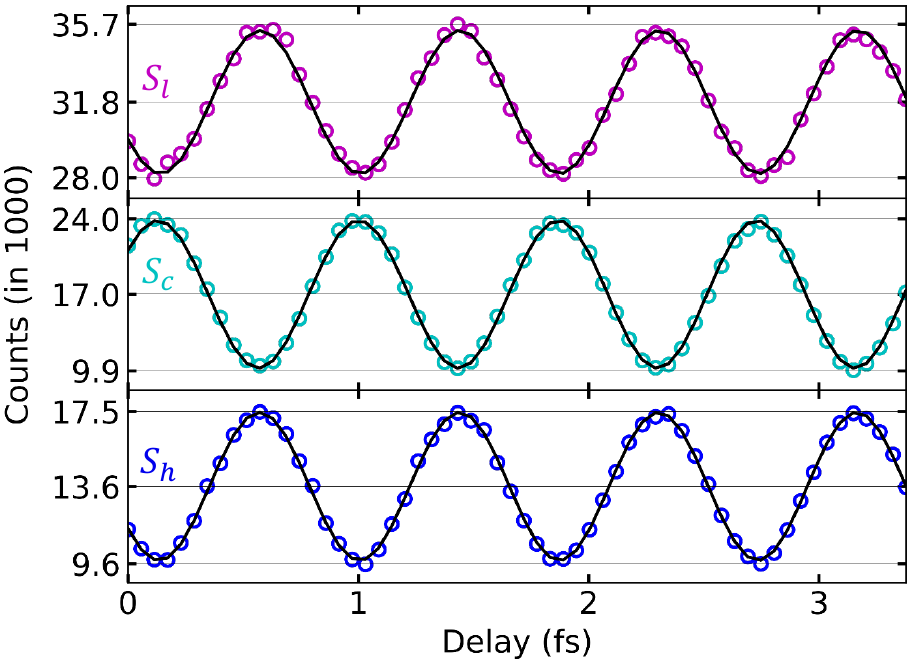} 
\caption{The delay-dependent photoelectron signal (dots) of the three sidebands in the $SB_{12}$ 
group and fits to a cosine function (lines).}
\label{fig:fit_12}
\end{figure}

Figure~\ref{fig:expt-results} exhibits the results of our 3-SB \hbox{RABBIT} experiment 
after integrating the signal over all photo\-electron emission angles. 
To highlight the oscillations, the \hbox{RABBIT} trace in panel~(a) is plotted after 
subtracting the average delay-integrated signal.  The delay-integrated photoelectron spectra 
(normalized to~1 at the highest peak) is plotted in panel~(b). 
Due to the high NIR intensity, some of the main bands are depleted substantially 
and appear weaker than the SBs in their vicinity. 
The angle-integrated photoelectron spectrum is integrated over a spectral window of 0.7 eV around the peak of the SBs. 

The \hbox{RABBIT} phase~($\phi_R$) is extracted by fitting a cosine function 
(cf.\ Eqs.~\eqref{eq: SB osc c})
to this delay-dependent oscillating signals of the sidebands, as seen in Fig.~\ref{fig:fit_12}. 
Due to the large dataset available and the excellent stability of the interferometer, 
the phase retrieval generally resulted in error bars 
smaller than the symbol size in Fig.~\ref{fig:expt-results}(c). 
This gives us confidence in the
results obtained from our extraction procedure.
The numerical values obtained for the various SB groups, as well the contrast ratio
\begin{equation}
\gamma\equiv \frac{{\rm max(SB}(\tau))-{\rm min(SB}(\tau))}{{\rm max(SB}(\tau))+{\rm min(SB}(\tau))} 
\end{equation}
are listed 
in Table~\ref{tab:Tab1}. As expected, the highest contrast is found 
for the center sideband, due
to the same ($3^{\rm rd}$) order of transitions involved.

We note that there are several auto\-ionizing resonances with principal configuration $3s3p^6 n\ell$ 
in the $SB_{12}$ range of photo\-electron kinetic energies, which converge towards the $(3s3p^6)^2S$
threshold of the first excited state of Ar$^+$ around 13.5~eV~\cite{NIST}.  Early measurements of the
$(3s3p^6 np)^1P^o$ resonances were reported by Madden {\it et al.}~\cite{PhysRev.177.136}.  They were also seen by Burke and Taylor~\cite{BurkeTaylor1975} in their photo\-ionization work, and further resonances with other configurations, which can be reached by charged-particle or multi\-photon impact, were discussed by Bartschat and Burke~\cite{BB88}.
More recently, the effect of these resonances on the RABBIT phase in \hbox{1-SB} setups was reported 
by Kotur {\it et al.}~\cite{Kotur2016} and Cirelli {\it et al.}~\cite{Cirelli2018}.  

Since we used the coupled-state
description of Burke and Taylor~\cite{BurkeTaylor1975}, we saw resonance effects 
in test calculations, but only with appropriate frequencies and sufficiently long pulses, 
for which the resonance widths could be well resolved.  
Note that these features are very sensitive 
to small fluctuations in the frequency and bandwidth of the APT during the XUV generation process. 
Therefore, these structures were not seen in the three experimental data points presented in the $SB_{12}$ region.
We hope to generate additional data with tunable high-order harmonic frequencies
in the future.  This will make it possible to investigate the resonance phenomena in more detail. 

As predicted by our generalized decomposition approximation~(cf.\ Eqs.~\eqref{eq: SB osc c}), 
the lower and the higher SBs oscillate by $\pi$ out of phase with the central SB.
The retrieved \hbox{RABBIT} phases $\phi_R$ are plotted in Fig.~\ref{fig:expt-results}(c)  
after removing the extra $\pi$ from $S_l$ and $S_h$ to simplify the comparison. 
The time-delay axis on the right side of this panel was created via the conversion $\tau_R=\phi_R/(4\omega)$. 

Five SB groups are clearly identifiable in Fig.~\ref{fig:expt-results}(c).
While there are some irregularities in $SB_{8}$ and $SB_{16}$, especially with the phase
extracted from $S_l$, groups $SB_{10}$, $SB_{12}$, and $SB_{14}$ show the expected trend:  
The RABBIT phases of the three SBs in each group are similar, although a small  
difference remains visible in $SB_{10}$.  That difference, however, essentially vanishes in $SB_{12}$ and $SB_{14}$.

The irregularity seen in the $SB_{8}$ group is due to a significant contribution of another 
4$^{\rm th}$-order transition in the absorption path of the lowest SB $S_l$, which involves a 
transition from M$_7$ down to the Rydberg states and back up to $S_l$. 
The Rydberg states enhance the strength of this transition and 
add a resonance phase that leads to 
a significant deviation in the RABBIT phase of $S_l$ compared to the other members of the $SB_{8}$ group. 
Furthermore, due to the low cut-off of the XUV spectrum based on HHG and the decreasing photo\-ionization 
cross section of argon with increasing photon energy, 
the strength of the M$_{17}$ peak is very weak compared to the rest of the lower main peaks. 
As a result, higher-order transitions involving lower main bands also play a significant 
role in the oscillation of $S_l$ in the $SB_{16}$ group, which again affects the extracted phase. 

\begin{figure*}
\includegraphics[width=0.90\linewidth]{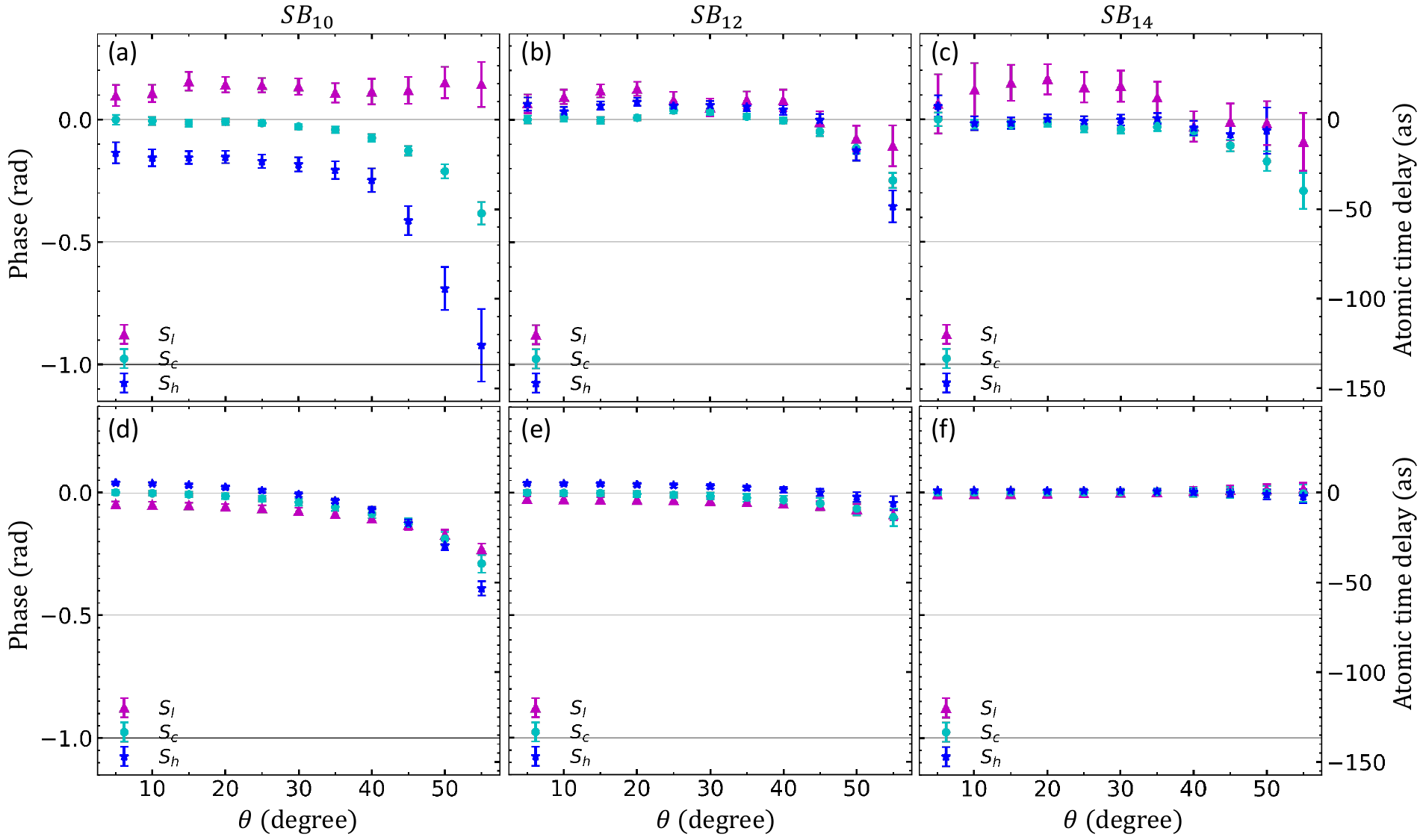} 
\caption{Top row: Angle-dependent \hbox{RABBIT} phases extracted from the measurements 
in group $SB_{10}$~(a), $SB_{12}$~(b), and $SB_{14}$~(c).  Bottom row: Corresponding RMT predictions.}
\label{fig:angle_exp_10_to_14}
\end{figure*}

\subsection{Angle-differential RABBIT phases}\label{subsec:Angle-differential}

We now further increase the level of detail by investigating angle-dependent RABBIT phases, which is possible due to the angle-resolving capability of the reaction microscope.
For the reasons given above regarding the additional complexities associated with the $SB_{8}$ and $SB_{16}$ groups,  we  concentrate the remaining discussion on $SB_{10}$, $SB_{12}$, and $SB_{14}$.

Figure~\ref{fig:angle_exp_10_to_14}(a-c) shows the \hbox{RABBIT} phases extracted within these groups as a function of the photo\-electron emission angle, which is defined relative to the (linear) laser polarization vector. 
The photoelectron signal is integrated over an angular window of $10^\circ$ for each data point.
The angle-resolved RABBIT phases are shifted to fix the starting phase of the central sideband in each group to zero.
According to both our experiment and the calculation (Fig.~\ref{fig:angle_exp_10_to_14}, panels \hbox{d-f}), the phase of $S_h$ exhibits a stronger angular dependence compared to that of $S_c$  and $S_l$.  
With increasing photo\-electron energy, the differences diminish in both experiment and theory, with theory predicting almost no angle-dependence in the range of $SB_{14}$ plotted.

To explain the angle-dependence in the RABBIT phase, we need to consider the interference among all the angular-momentum channels of the sidebands accessed through the absorption and emission paths.
We write the signal in compressed form as 
\begin{align}
S_{q}(\tau,\theta)  &\propto   \sum_{\ell,\ell',m} \alpha^a_{\ell,m}\alpha^{e^*}_{\ell',m}  Y_{\ell,m}(\theta)Y^*_{\ell',m}(\theta) \nonumber\\  &  \quad \quad \quad \times \cos(4\,\omega\tau -\Delta\phi_{\Omega}^{q}-\Delta\phi_{\ell\ell',m}^{at})  \nonumber\\ 
&\propto  I_1(\theta) \, \cos(4\,\omega\tau -\phi_{R}(\theta)).
\end{align}
\label{eq: SB theta}
\noindent Here $\alpha^{a}$ and $\alpha^{e}$ are the transition amplitudes involving the various fields and matrix elements, while $\ell (\ell')$ denotes the angular-momentum channels accessed through the absorption (emission) path.  

The dissimilarity in the RABBIT phases ($\phi_{R}(\theta)$) of the three SBs can be explained by considering a propensity rule for the transition amplitudes and  the dependence of both the Wigner and $\phi^{cc}$ phases on the orbital angular momenta. 
It is well known that the Wigner phase depends on the angular momentum channel.  
The cc phase has also been shown to depend slightly on whether there is an increase or decrease in the angular momentum, while it appears to remain independent of the target species~\cite{Fuchs:s,Peschel2022}.
Therefore, the atomic phases $(\Delta\phi_{\ell\ell',m}^{at})$ arising from the interference between various \hbox{$\ell$-channels} of emission and absorption paths are also expected to differ. 
Similar to bound-continuum transitions~\cite{PhysRevA.32.617}, absorption (emission) within the continuum favors an increase (decrease) in the angular momentum of the outgoing photo\-electron, especially for low kinetic energies~\cite{Bertolino_2020,PhysRevLett.123.133201,Peschel2022, Kheifets2022,PhysRevA.106.023116}. 
The higher SB ($S_h$) of the group involves the absorption of three probe photons \hbox{($H_{q-1}+3\,\omega$)} that, according to the propensity rule, predominantly populate higher angular-momentum states. 
Along the other path \hbox{($H_{q+1}-1\,\omega$)} leading to~$S_h$, the emission of one probe photon mainly creates lower angular-momentum states.  
For $S_l$, emission of three probe photons \hbox{($H_{q+1}-3\,\omega$)} primarily leads to the population of lower angular-momentum states.
Even though the absorption path \hbox{($H_{q-1}+1\,\omega$)} to $S_l$ also favors an increase in the photoelectron's angular momentum, the possible values reached by the absorption of a single probe photon remain relatively small. 

The interplay of the propensity rule for transition amplitudes to each $\ell$-channel and the angle-dependent amplitudes of the coupled spherical harmonics determine the angular variation of $\phi_{R}$ in the three SBs. 
In cross-channel interference, $\ell \neq \ell'$, the angle-dependent spherical harmonics undergo a sign change across their angular nodes, thus resulting in a phase jump by~$\pi$.
If the relative magnitude of these cross-channel interferences is significant compared to that of the same-channel interference terms, \hbox{$\ell = \ell'$}, this can lead to a rapid variation in the angle-dependence of $\phi_R$ in the vicinity of the nodes~\cite{PhysRevLett.123.133201}. 
Depending on the value of $\Delta\phi_{\ell\ell',m}^{at}$ relative to the average $\Delta\phi^{at}$ of the interference terms, the additional $\pi$-jump at the nodes in $Y_{\ell,m}$ and/or $Y_{\ell',m}$ can drive the angle-dependent curve downward or upward. 

With increasing $\ell$ value, the position of the first node in the associated Legendre polynomial of the spherical harmonic moves to smaller angles. 
Due to the propensity rule, the weight of the cross-channel interference term containing large $\ell$-values is most significant in the higher sideband. 
This results in a relatively early onset of the descent in the angle-dependent RABBIT phase in the higher sideband. In the lower sideband, the amplitude of the cross-channel interference term containing large $\ell$-values is not very strong; hence, the $\pi$-jump across the node does not produce a substantial change in the overall retrieved phase.   
With increasing kinetic energy, for both the absorption or emission of the probe photons, the  transition amplitudes for increasing and decreasing angular momentum tend to become similar \cite{PhysRevLett.123.133201}.  
Hence the contribution of cross-channel interference containing large $\ell$ values decreases with increasing kinetic energy. Thus the $\pi$-jumps at the nodes of the corresponding spherical harmonics do not change the retrieved phase significantly.

Since the retrieved angle-integrated RABBIT phase is the weighted average of all the channel-resolved RABBIT phases and the weights of these channels in the $S_l$, $S_c$, and $S_h$ SBs are different, the angle-integrated RABBIT phase in the three SBs also turns out different. 
Also, owing to the propensity rule, the unequal transition probabilities of reaching the various angular momentum states of the SBs in absorption and emission of the probe photons may cause incomplete interference in the individual $\ell$ channels, thereby reducing the overall oscillation contrast in the angle-integrated photoelectron signal.

Finally, we notice that the scale of the variations in the angle-dependence of the RABBIT phase 
depicted in Fig.~\ref{fig:angle_exp_10_to_14} is 
smaller in the calculation than in our experiment. 
Also, the positions of $S_l$ 
and $S_h$ relative to $S_c$ appear to be switched. 
In addition to always possible shortcomings in the theoretical model (as sophisticated as it might be)
and unknown potential systematic errors in the experiment, 
the difference in the probe intensities and the pulse details, in general, are likely responsible for at least some of the discrepancies seen here. 
We hope to be able to investigate this in more detail in the future by performing additional
calculations with different intensities and more time delays.

\section{Summary and Outlook}\label{sec:outlook}
\smallskip
In summary, we carried out a proof-of-principle 3-SB \hbox{RABBIT} experiment in argon.
In contrast to more popular single-SB studies, our technique enables us to focus on 
the photon-induced transition phases without distortion from a possibly unknown 
or experimentally drifting XUV chirp. 
While we confirmed earlier predictions that the angle-integrated \hbox{RABBIT} phases 
extracted within a SB group become increasingly similar, we enhanced the analyzing 
power of the setup significantly by resolving the emission angle with a reaction microscope.  
By doing so, we could identify which of the three sideband phases within a group is 
most sensitive to a change in the detection angle.  

Our experimental efforts were accompanied by numerical calculations performed with the 
non\-perturbative all-electron \hbox{$R$-matrix} with time dependence method.  
There is some qualitative agreement between experiment and theory regarding the general 
trends observed,  but significant differences remain in the details.  
Given the remaining limitations and challenges faced in the present study, especially concerning the details of 
the pulse and the argon target, the remaining deviations between experiment and theory 
in the quantitative values of the phases are not too surprising.
We hope to address these issues in future improvements of the setup.

As the next step, we plan to repeat this experiment with helium, where the contribution of the Wigner phase 
for an $s \to p$ transition remains the same in all three sidebands. Any differences in the phases within the group 
then clearly indicate the influence of $\phi^{cc}$. This switch of targets would require 
extending the harmonic cut-off, 
which is by no means trivial in our scheme, as the cut-off in the HHG process 
decreases with the driving frequency. Using helium instead of argon also has the advantage of
theory likely being more reliable due to the simplicity of the target.  On the other hand,
heavier quasi-two-electron targets with an $(ns^2)^1S$ outer-shell configuration 
(unfortunately, these are metals that would need to be vaporized rather than inert gases) would 
provide a larger short-range modification of the relevant interaction potential and, therefore,
may be more suitable to investigate whether $\phi^{cc}$ is indeed nearly universal.

Undoubtedly, many open questions will need to be answered before the effect of the 
additional continuum-continuum transitions in single- and multiple-SB \hbox{RABBIT} setups are fully understood. 
It would be interesting to analyze whether the SB phases always converge to each other with increasing energy, 
whether or not they cross in a predictable way with increasing emission angle,
and how the behavior depends on the target investigated.
While we cannot answer these questions at the present time, we hope that other groups will see 
the work reported in this paper 
as a worthwhile inspiration to carry out further studies in this field.

\medskip
\begin{acknowledgments}
The experimental part of this work was supported by the DFG-QUTIF program under
Project \hbox{No.~HA 8399/2-1} and \hbox{IMPRS-QD}.
K.R.H.\ and K.B.\ acknowledge funding from the NSF through grants
\hbox{No.~PHY-1803844} and \hbox{No.~PHY-2110023}, respectively, as well as the  
Frontera Pathways allocation PHY-20028.
\end{acknowledgments}
\bibliographystyle{apsrev4-1}

\end{document}